\documentclass[a4paper, 10pt, conference]{IEEEtran}
\ifCLASSOPTIONcompsoc
  \usepackage[caption=false,font=normalsize,labelfont=sf,textfont=sf]{subfig}
\else
  \usepackage[caption=false,font=footnotesize]{subfig}
\fi
\usepackage{graphicx}
\IEEEoverridecommandlockouts
\usepackage{cite}
\usepackage{amsmath,amssymb,amsfonts}
\usepackage{algorithmic}
\usepackage{graphicx}
\usepackage{subfig}
\usepackage{textcomp}
\usepackage{xcolor}
\def\BibTeX{{\rm B\kern-.05em{\sc i\kern-.025em b}\kern-.08em
    T\kern-.1667em\lower.7ex\hbox{E}\kern-.125emX}}
\begin{document}

\title{Enhanced Synthetic MRI Generation from CT Scans Using CycleGAN with Feature Extraction}

\author{
\IEEEauthorblockN{
Saba Nikbakhsh\IEEEauthorrefmark{1}, 
Lachin Naghashyar\IEEEauthorrefmark{2},
Morteza Valizadeh\IEEEauthorrefmark{1}, 
Mehdi Chehel Amirani\IEEEauthorrefmark{1}, 
}
\IEEEauthorblockA{\IEEEauthorrefmark{1}Department of Electrical Engineering\\
Urmia University of Technology\\
Urmia, Iran\\
Email: sabanikbakhsh@gmail.com, mo.validzadeh@urmia.ac.ir, m.amirani@urmia.ac.ir}
\IEEEauthorblockA{\IEEEauthorrefmark{2}Department of Computer Science\\
Sharif University of Technology\\
Tehran, Iran\\
Email: lachin.naghashyar@sharif.edu}
}

\maketitle

\begin{abstract}
In the field of radiotherapy, accurate imaging and image registration are of utmost importance for precise treatment planning. Magnetic Resonance Imaging (MRI) offers detailed imaging without being invasive and excels in soft-tissue contrast, making it a preferred modality for radiotherapy planning. However, the high cost of MRI, longer acquisition time, and certain health considerations for patients pose challenges. Conversely, Computed Tomography (CT) scans offer a quicker and less expensive imaging solution. To bridge these modalities and address multimodal alignment challenges, we introduce an approach for enhanced monomodal registration using synthetic MRI images. Utilizing unpaired data, this paper proposes a novel method to produce these synthetic MRI images from CT scans, leveraging CycleGANs and feature extractors. By building upon the foundational work on Cycle-Consistent Adversarial Networks and incorporating advancements from related literature, our methodology shows promising results, outperforming several state-of-the-art methods. The efficacy of our approach is validated by multiple comparison metrics.

\end{abstract}

\section{Introduction}

Medical imaging has revolutionized healthcare, offering clinicians and researchers deep insights into the human body. Among the various applications, radiotherapy is a standard treatment method for many diseases. However, precisely positioning the patient to target unhealthy tissue is one of the main challenges. This positioning integrates information from hard tissues, as seen in Computed Tomography (CT) scans, and soft tissues, predominantly captured by Magnetic Resonance Imaging (MRI), a process known as medical image registration. However, the inherent differences in image quality, contrast, and representation between MRI and CT can introduce discrepancies, potentially leading to treatment errors.

MRI, while superior in capturing soft tissue details, is time-consuming (taking up to five times longer than CT), more costly, and poses challenges for patients with metallic implants, such as metal rods in joints, pacemakers, or those suffering from claustrophobia. Despite these challenges, the exceptional imaging capabilities of MRI, especially for soft tissues, have ignited significant interest in synthesizing MRI images from CT scans.

Recent advances in deep learning, especially in medical computer-aided diagnosis, have shown remarkable performance in applications of medical images such as image synthesis. Previously, many methods mainly focused on synthesizing CT images from existing MR images for dose planning and MR-only radiotherapy \cite{Liu2020,Dinkla2019,BrouBoni2020,Han2017,Chen2022,Florkow2022,Kazemifar2019,Lei2019,Liu2019,Qi2020}. Meanwhile, only a few studies have focused on CT-to-MRI synthesis.

Traditional methods often employed Convolutional Neural Networks (CNNs) \cite{Xiao2017}, but their computational intensity led researchers like Zhao et al. \cite{Zhao2017} to propose a UNet structure for CT to MRI conversion. However, CNN and UNet-based methods, relying on voxel-wise loss functions, tend to produce blurred images. To overcome this problem, Nie et al. \cite{Nie2017} combined voxel-wise loss with adversarial loss, laying the foundation for developing Generative Adversarial Networks (GANs). This approach was further adopted and refined by researchers like Hong et al. \cite{Hong2022}. Following studies combined GANs with other architectures, as seen in works by Li et al. \cite{Li2020}, Dai et al. \cite{Dai2021}, and Feng et al \cite{Feng2022}. Isola et al. \cite{Isola2017} introduced the Pix2Pix network, a supervised method that operates in a pixel-to-pixel manner, underscoring the importance of paired data for its operation. The introduction of CycleGAN by Wolterink et al.\cite{Wolterink2017} further eliminated the absolute need for paired data, and many researchers have since leveraged this network for MRI image synthesis. In pelvic studies, Dong et al. \cite{Dong2019} utilized the cycle-consistent deep attention network to generate MRI images and then segment them. Meanwhile, Kalantar et al.\cite{Kalantar2021} employed UNet and CycleGAN for CT-based pelvic MR synthesis, noting challenges with CycleGAN in handling soft-tissue misalignments.

In this study, our primary contribution is introducing a novel approach that seamlessly integrates CycleGANs with a feature extractor architecture. This combination seeks to utilize the power of CycleGANs in handling unpaired data. At the same time, the feature extractor is designed to capture and retain fine details, often lost in the conventional synthesis process. By doing so, our method addresses the issue of data pairing and ensures that the synthesized MRI images retain a higher fidelity than their CT counterparts. The results are assessed against other models, including Pix2Pix, U-Net, and simple CycleGAN methods, demonstrating superior accuracy and quality.

\section{Materials and Method}

\subsection{Dataset}

The study employed the dataset from the ``Gold Atlas'' project \cite{Nyholm2018}, which includes paired MRI and CT scans of the pelvic region from 19 male patients. These images are provided in the DICOM format, with each set representing a three-dimensional structure constructed from multiple two-dimensional axial slices in grayscale. The data preprocessing steps encompassed: 

\begin{enumerate}
    \item \textbf{Normalization:} Due to the variability in pixel values across different imaging modalities, standardizing the dataset was crucial for consistent processing. As a result, all images were converted to floating-point representations within the $[0, 255]$ grayscale range and then mapped to the $[-1, 1]$ range.
    
    \item \textbf{Augmentation:} The original images, with dimensions of $512\times512$ pixels, were resized to $256\times256$. Following data augmentation and the exclusion of unclear and noisy images, The final dataset contains $2884$ images of CT and MRI T2, which are matched pixel by pixel and are so-called pairs. The applied augmentation techniques aimed to enhance the model’s robustness and generalization capabilities:
    \begin{itemize}
        \item \textbf{Random Crop:} The images were initially expanded to $572\times572$ pixels, then randomly cropped to $512\times512$, and finally resized to $256\times256$ pixels.
        \item \textbf{Horizontal Flip:} The cropped CT and MR images had a $50$\% probability of undergoing horizontal flipping.
        \item \textbf{Rotation:} To account for variations in patient orientation, the flipped images were randomly rotated between $-5$ and $5$ degrees.
    \end{itemize}
\end{enumerate}

\subsection{Model Architecture}

Our proposed model, CycleGAN+FE (CycleGAN Enhanced with a Feature Extractor), extends the traditional CycleGAN architecture by integrating a feature extractor to enhance image translation performance. This model employs CycleGAN, a specialized form of Generative Adversarial Network. This model's main role is to transform images from their initial domain into the designated target domain. Notably, this process does not require a paired dataset, a characteristic known as unpaired image-to-image translation. This approach aligns with the architecture described in \cite{Zhu2017}. This network uses a forward and a backward cycle to translate images. In the forward cycle, the primary generator \(G_{MR}\) aims to convert a CT image into its MRI equivalent. However, one of the common problems in the CycleGAN network is mode collapse, where the network generates only a small set of images or a duplicate image capable of deceiving the discriminator network. To mitigate the mode collapse issue, the synthesized MRI image is subsequently input into a second generator, aiming to reconstruct the original CT image, a process termed cycle consistency. Concurrently, the discriminator evaluates the authenticity of the generated MRI image, distinguishing between real and synthetic outputs.

A parallel backward cycle aims at CT synthesis from MRI inputs using the generator \(G_{CT}\). 

The cycle consistency helps prevent mode collapse and unnecessary detail generation by the generators. 
However, as highlighted in specific studies \cite{Na2019} and \cite{Wang2018}, it occasionally introduces ambiguity within the network, mainly when translating from synthetic MRI images and attempting to reconstruct authentic CT images.

To address the challenges in image translation, the Researchers offer a variety of solutions. One method involves utilizing mutual information loss for CT image generation \cite{mutualinfo}. Nevertheless, MRI images typically contain more intricate details. A notable approach, as discussed in \cite{Wang2018}, is the implementation of a feature extractor. This is achieved by omitting the final layer of the discriminator. However, the precise architecture of the feature extractor, its associated loss functions, and other configuration parameters remain crucial considerations, especially in medical imaging, as they can significantly influence the network's accuracy. Notably, the design of the feature extractor aligns with the $70 \times 70$ patch GAN discriminator structure \cite{Zhu2017}.

\begin{figure}[!t]
    \centering
    \includegraphics[width=1\linewidth]{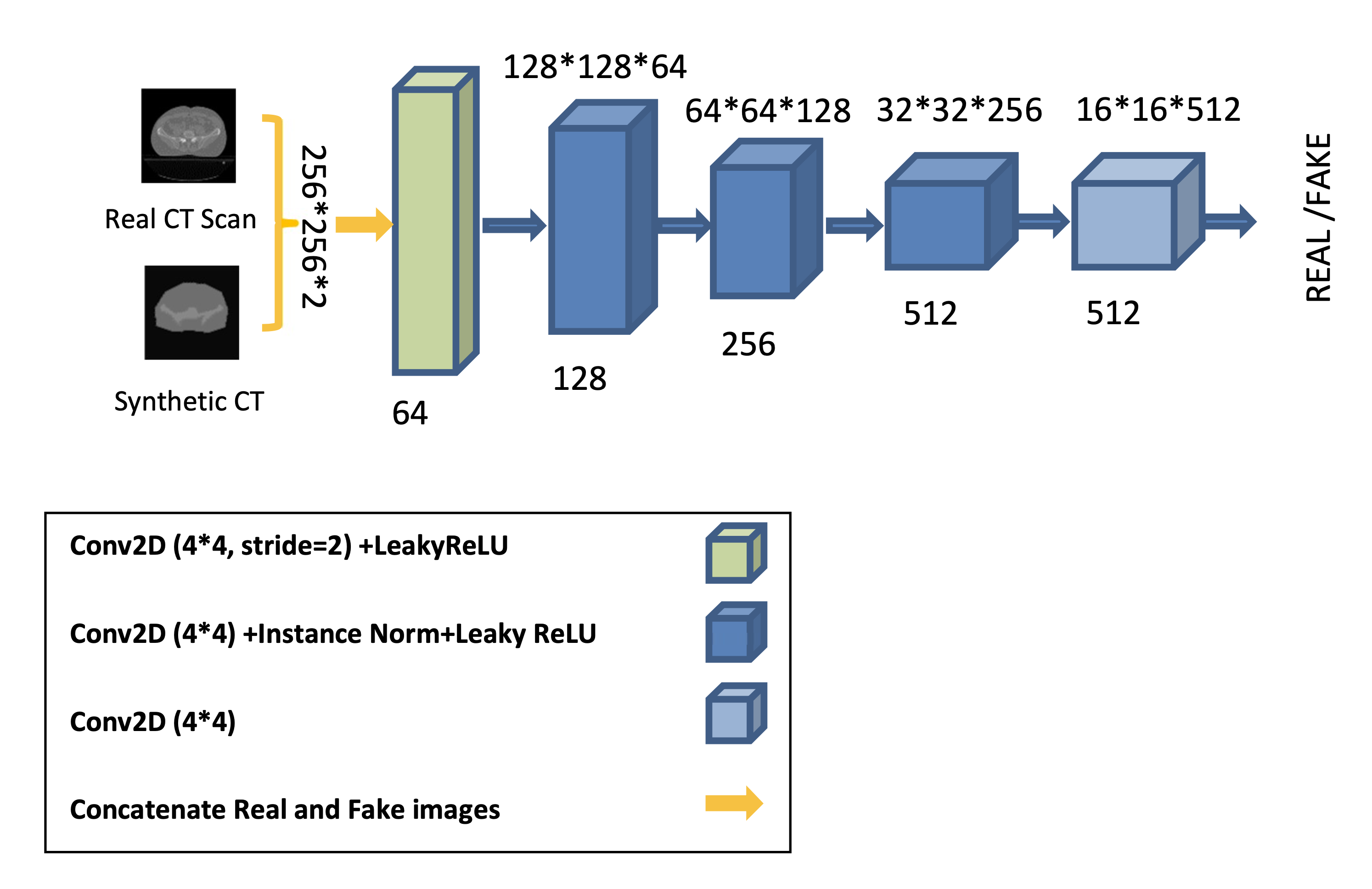}
    \caption{Architecture of the Feature Extractor}
    \label{fig:feature_extractor}
\end{figure}

In our network, the patch GAN discriminator's last layer is excluded for use as a feature extractor. Figure~\ref{fig:feature_extractor} presents a detailed illustration of this structure. This feature extractor plays a pivotal role in enhancing the performance of our model by ensuring that the most relevant features are captured and utilized during the image translation process. 

\begin{figure}[!t]
    \centering
    \includegraphics[width=1\linewidth]{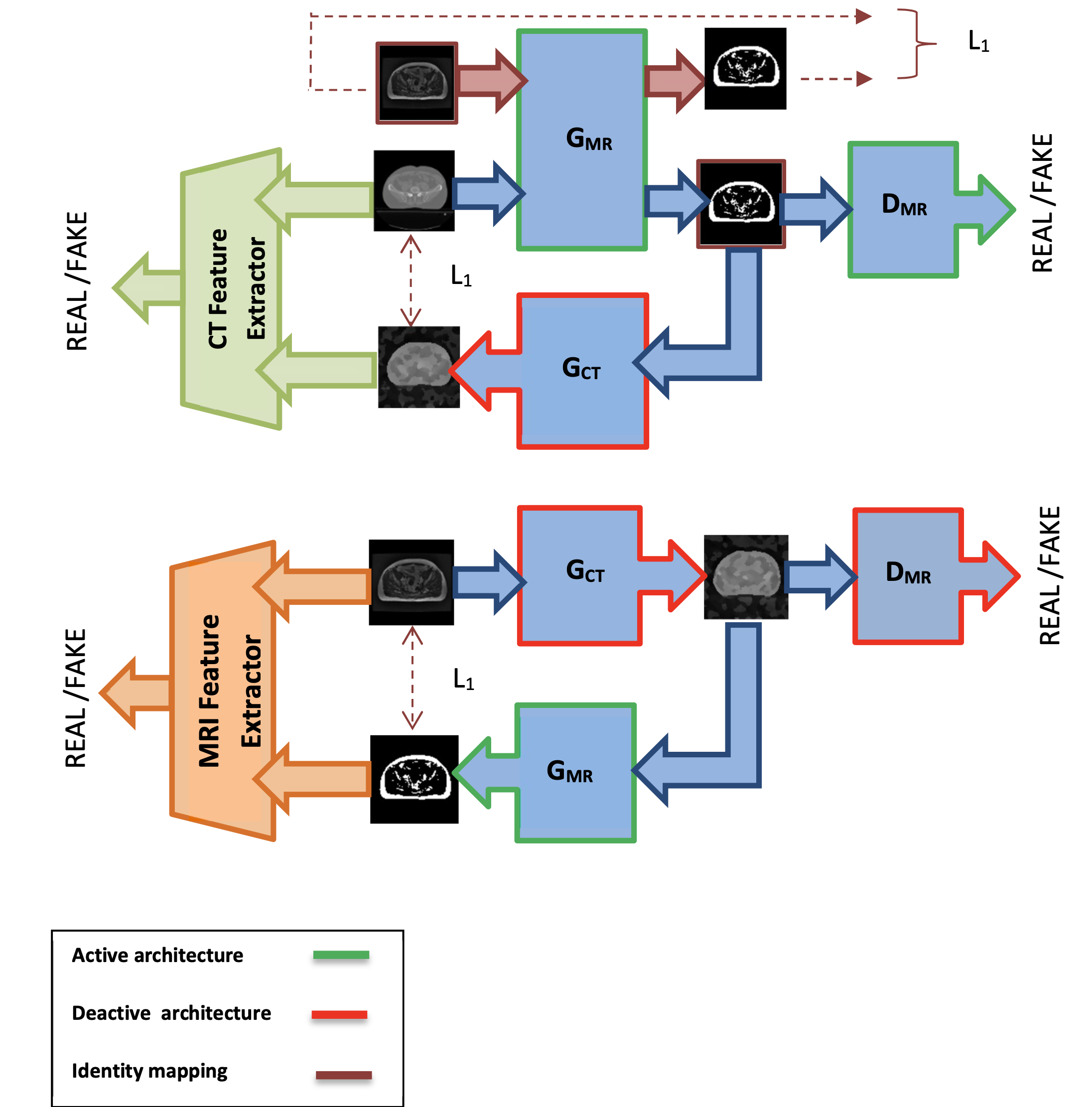}
    \caption{Proposed CycleGAN architecture for updating \(G_{MR}\) weights}
    \label{fig:my_label}
\end{figure}

Regarding model configuration and training, our model seamlessly integrates the generator, discriminator, and feature extractor components. This ensures synchronized training while allowing individual updates for each module. The training process is split into phases A and B that focus on the weights of \(G_{MR}\) and \(G_{CT}\)  to facilitate the translation of CT images to MRI ones and vice versa, respectively. Figure~\ref{fig:my_label} visually depicts the network structure for the CT scan to MRI translation. Specifically:
\begin{itemize}
    \item \(G_{MR}\), depicted in green, is updated during the phase A.
    \item In contrast, \(G_{CT}\), represented in red, generates images using its initial weights and remains unchanged during phase A.
\end{itemize}

A similar process occurs when updating the weights of \( G_{CT} \) for CT synthesis. During phase B, \( G_{CT} \) undergoes weight updates while the weights of \( G_{MR} \) remain unchanged.

\subsection{Loss Functions}

Loss functions play a fundamental role in the refinement of network parameters. In the rest of this section, we will explain various models and associated loss functions in detail, starting with \(G_{MR}\). The following subsections describe the specific loss functions and their importance in refining \(G_{MR}\). Later in the section, we also cover the updates to the parameters of \(D_{MR}\), the CT feature extractor, and similarly, \(G_{CT}\), \(D_{CT}\), and the MR feature extractor.

\begin{enumerate}
    \item \textbf{Adversarial Loss:} 
    After initializing the network parameters, a random CT scan image is selected and input into the \(G_{MR}\) generator. Using its initial parameters, the generator outputs a synthetic MRI image. This image is then evaluated by the $D_{MR}$ discriminator for authenticity. Typically, genuine images are labeled as one, while synthetic ones are labeled as zero. The discriminator's output lies in the range of $[0,1]$. The adversarial loss, described in Equation~\ref{eq:adversarial_loss}, measures the output deviation from the ideal label of one.
    
    \begin{align}
    \label{eq:adversarial_loss}
    & L_{GAN}(\operatorname{G}_{MR}, D_{MR}, I_{CT}, I_{MR}) = \nonumber \\
    & E_{I_{CT} \sim P_{\operatorname{data}}(I_{CT})}[\operatorname{D}_{MR}(\operatorname{G}_{MR}(I_{CT}))-1]^2
    \end{align}

    \item \textbf{Identity Loss:}
    Our network utilizes an identity mapping strategy. This approach guarantees that when an image from the target dataset, in this instance, the MRI dataset, is input into the corresponding generator, the resulting output closely mirrors the input. This method effectively maintains the intrinsic properties of the original image. In mathematical terms, the difference between the original MRI and the output from \(G_{MR}\) is determined as per Equation 2.

    \begin{align}
    \label{eq:identity}
    & L_{\text{identity}}(\operatorname{G}_{MR}, I_{MR}) = \nonumber \\ & E_{I_{MR} \sim P_{\text{data}}(I_{MR})} \left[ \left\| \operatorname{G}_{MR}(I_{MR}) - I_{MR} \right\|_1 \right]
    \end{align}

    \item \textbf{Cycle Consistency Loss:} 
    The generated MRI image is then input into the $G_{CT}$ generator, aiming to reconstruct the original CT scan. The difference between the genuine and reconstructed CT scan images is computed per Equation~\ref{eq:cycle_consistency}. In a reverse process, a real MRI image is input to \(G_{CT}\), resulting in a synthetic CT scan image. This image is then processed by \(G_{MR}\) for reconstruction, with the associated loss being determined by the provided formula, where \(I_{CT}\) and \(I_{MR}\) are real images, and \(G_{CT}(G_{MR}(I_{CT}))\) and \(G_{MR}(G_{CT}(I_{MR}))\) are reconstructed images.

    \begin{align}
    \label{eq:cycle_consistency}
    & L_{\text{CC}}(\operatorname{G}_{MR}, \operatorname{G}_{CT})  =  \nonumber \\
    & E_{I_{CT} \sim P_{\text{data}}(I_{CT})} [ \lVert \operatorname{G}_{CT}(\operatorname{G}_{MR}(I_{CT})) - I_{CT} \rVert_1 ] \nonumber \\
    & +  E_{I_{MR} \sim P_{\text{data}}(I_{MR})} [ \lVert \operatorname{G}_{MR}(\operatorname{G}_{CT}(I_{MR})) - I_{MR} \rVert_1 ]
    \end{align}

    \item \textbf{Feature Vector Loss:}

The Feature Extractor network processes original and reconstructed CT scan images to derive feature vectors. These vectors are essential metrics for determining image authenticity. To ensure the accuracy of these vectors, we used the Mean Squared Error (MSE) loss. The reason for choosing MSE is its ability to focus on the details of images, which is vital for medical imaging applications. Through testing, we found that MSE provided better results than Mean Absolute Error (MAE), particularly in preserving the intricate details of the images.

As shown in Figure~\ref{fig:feature_extractor}, the Feature Extractor network processes both real and synthetic CT images. It employs various layers to process the images progressively. The ultimate goal of the network is to classify the image as either real or synthetic.

In terms of the mathematical representation, when \( \Psi_{CT} \) represents the transformation of the real CT scan image through the generator network, i.e., \( \Psi_{CT} = G_{CT}(G_{MR}(I_{CT})) \), then \( \Psi_{CT} \) is concatenated with \( I_{CT} \) and passed through the feature extractor. Similarly, \( \Psi_{MR}  = G_{MR}(G_{CT}(I_{MR}))\). The mean squared error between the feature vectors is calculated using Equation~\ref{eq:feature_ext}. 

\begin{align}
\label{eq:feature_ext}
& L_{\text{FE}}(\operatorname{G}_{MR}, \operatorname{G}_{CT}, D_{MR}, D_{CT}, I_{CT}, I_{MR}) =  \IEEEnonumber\\
& E_{I_{CT} \sim P_{\text{data}}(I_{CT})} [ (f_{D_{CT}}(\Psi_{CT}, I_{CT}) - 1)^2] \IEEEnonumber\\ 
& + E_{I_{MR} \sim P_{\text{data}}(I_{MR})} [ (f_{D_{MR}}(\Psi_{MR}, I_{MR}) - 1)^2 ]
\end{align}

\end{enumerate}

Based on Equation~\ref{eq:gan_loss}, the overall loss function is derived by combining various loss functions with specific coefficients. These coefficients are determined based on prior research and empirical observations. The individual coefficients are as follows:

\begin{itemize}
    \item \textbf{Adversarial Loss Coefficient:} Typically set to $1$, it determines the weightage of the adversarial loss in the overall loss computation.
    \item \textbf{Identity Loss Coefficient (\(\alpha\)):} Set to $5$, this coefficient emphasizes the importance of retaining the original image characteristics during transformations.
    \item \textbf{Cycle Consistency Loss Coefficient (\(\beta\)):} For our implementation, \(\beta\) is set to $10 - \gamma = 9.99$, indicating its significance in ensuring the reconstructed image closely resembles the original.
    \item \textbf{Feature Extractor Loss Coefficient (\(\gamma\)):} Given that the mean squared error (MSE) is squared, it inherently has a stronger impact. Therefore, beginning with a smaller coefficient during the initial epochs is better, ideally within the $[0,1]$ range. Through experimental evaluations, we set \(\gamma\) to $0.01$.

\end{itemize}

Adopting these coefficient values ensures that the generator is not biased towards merely replicating the input image, especially in later epochs. This cumulative loss then undergoes backpropagation, prompting updates to the \(G_{MR}\) parameters.

\begin{align}
\label{eq:gan_loss}
& L_{G_{\text{MR}}} = L_{GAN}(\operatorname{G_{MR}}, D_{MR}, I_{CT}, I_{MR}) \IEEEnonumber \\
&\quad + \alpha L_{\text{identity}}(\operatorname{G_{MR}}, I_{MR}) \IEEEnonumber \\
&\quad + \beta L_{\text{CC}}(\operatorname{G_{MR}}, \operatorname{G_{CT}}) \IEEEnonumber \\
&\quad + \gamma L_{\text{FE}} (\operatorname{G_{MR}}, \operatorname{G_{CT}}, D_{CT}, D_{MR}, I_{CT}, I_{MR}) 
\end{align}

In Phase A, following the update of the generator \(G_{MR}\)'s weights, we proceed to adjust the parameters of \(D_{MR}\) and the CT feature extractor. The discriminator \(D_{MR}\) measures the similarity between the synthetic MR images created by \(G_{MR}\) and real MR images, with its refinements and loss function detailed in Equation \ref{eq:discriminator_loss}. On the other hand, the Feature Extractor for CT images is designed to capture the essential features of the CT images, with its loss function and refinements specified by Equation \ref{eq:f_ex_loss}.

\begin{align}
\label{eq:discriminator_loss}
& L_{GAN_{\text{update}}}(\operatorname{G}_{MR}, \operatorname{D}_{MR}, I_{CT}, I_{MR}) = \IEEEnonumber \\
& E_{I_{MR} \sim P \operatorname{data}(I_{MR})}[\operatorname{D}_{MR}(I_{MR})-1]^2 \IEEEnonumber \\
& + E_{I_{CT} \sim P \operatorname{data}(I_{CT})}[\operatorname{D}_{MR}(\operatorname{G}_{MR}(I_{CT}))^2]
\end{align}

\begin{align}
\label{eq:f_ex_loss}
& L_{FE_{\text{update}}}(\operatorname{G}_{MR}, \operatorname{G}_{CT}, D_{CT}, I_{CT}) = \IEEEnonumber \\
& \quad E_{I_{CT} \sim P_{\text{data}}(I_{CT})} [f_{D_{CT}}(\Psi_{CT}, I_{CT})]^2 \IEEEnonumber \\
& + E_{I_{CT} \sim P \text{data}(I_{CT})} [f_{D_{CT}}(I_{CT}, I_{CT})-1]^2
\end{align}

In phase B, \(G_{MR}\) stops updating its parameters, while \(G_{CT}\), \(D_{CT}\) and the MRI feature extractor adjust and update their weights, following the similar procedure and loss criteria.

\subsection{Training Details}

 We have used the Nvidia RTX 3070 GPU to train our model. The network is trained on 2884 unpair CT and MR images in each epoch randomly selected from the pair dataset. The batch size is set to one, implying that a single CT scan or MRI image is processed. The network's error in synthesizing MRI from CT scans undergoes backpropagation to refine the weights. We used the image pool technique to stabilize the training process, which stores the last 50 synthesized MRI images. In each iteration, one of these images is randomly selected. This approach ensures that most synthesized images are utilized multiple times, leading to a more stable loss rate. The training consisted of 200 epochs. The learning rate was set to $0.0002$ for half of the epochs, which decreased linearly for the subsequent $100$ epochs. The Adam optimizer was employed, consistent with standard deep learning practices.

\section{Results}

Several supervised and unsupervised networks were employed to reconstruct MRI images from CT scans. Their performance was evaluated using 235 test CT and MR images that were not used during the training phase. Quantitative and qualitative metrics such as MAE, MSE, PSNR, and SSIM were utilized to evaluate the fidelity of translated MR images to the real ones.

To provide clarity:
\begin{itemize}
    \item \textbf{Mean Absolute Error (MAE)}: Calculated as:
    \[
    \text{MAE} = \frac{1}{N} \sum_{i=1}^{N} |R_i - S_i|
    \]
    where \(R_i\) is the real CT or MR image pixel and \(S_i\) is the synthetic one. $N$ is also the total number of pixels.

    \item \textbf{Mean Squared Error (MSE)}: Calculated as:
    \[
    \text{MSE} = \frac{1}{N} \sum_{i=1}^{N} (R_i - S_i)^2
    \]

    \item \textbf{Peak Signal-to-Noise Ratio (PSNR)}: A quality metric defined as: 
    \[
    \text{PSNR} = 20 \times \log_{10}\left(\frac{\text{MAX}_I}{\sqrt{\text{MSE}}}\right)
    \]
    In this context, \(\text{MAX}_I\) is the highest possible pixel value in the image.

    \item \textbf{SSIM (Structural Similarity Index)}: Measures the similarity between real and synthetic MR images:
    \[
    \text{SSIM}(I_R, I_S) = \frac{(2\mu_{I_{R}}\mu_{I_{S}} + c_1)(2\sigma_{{I_{R}}{I_{S}}} + c_2)}{(\mu_{I_{R}}^2 + \mu_{I_{S}}^2 + c_1)(\sigma_{I_{R}}^2 + \sigma_{I_{S}}^2 + c_2)}
    \]
    where \(\mu_{I_{R}}\) and \(\mu_{I_{S}}\) are the averages of real and synthetic images, and \(\sigma_{{I_{R}}{I_{S}}}\) is their covariance.
\end{itemize}

The CycleGAN+FE, a CycleGAN enhanced with a feature extractor, exhibited superior performance throughout the training. Notably, it stood out when trained with Pair data, reducing error rates and improving image clarity. We chose to evaluate the Pair variant of the CycleGAN to facilitate a direct comparison with other supervised methods, and it not only aligned with but also outperformed these paradigms.

In contrast, supervised networks like Pix2Pix \cite{Isola2017} and UNet \cite{Li2020} (with L1 loss) showed an early stabilization post 50 epochs with limited subsequent enhancements. The relevant results based on the evaluation metrics are shown in Table \ref{tab:model_comparison}.

\begin{table}[!t]
\caption{Comparative Analysis of Models Based on Quantitative Metrics}
\begin{center}
\begin{tabular}{|c|c|c|c|c|}
\hline
NETWORKS & MAE & MSE & PSNR & SSIM \\
\hline
UNet & 14.43 & 770.30 & 20.19 & 0.66 \\
\hline
Pix2Pix & 13.15 & 657.55 & 20.83 & 0.66 \\
\hline
CycleGAN Unpair & 13.09 & 648.94 & 20.7 & 0.62 \\
\hline
CycleGAN+FE Unpair & 13.2 & 631.51 & 20.85 & 0.62 \\
\hline
CycleGAN+FE Pair & 12.26 & 549.19 & 21.43 & 0.62 \\
\hline
\end{tabular}
\label{tab:model_comparison}
\end{center}
\end{table}

\subsection{Evaluation of Results Based on Qualitative Criteria}

\begin{figure}[!t]
    \centering
    \includegraphics[width=1\linewidth]{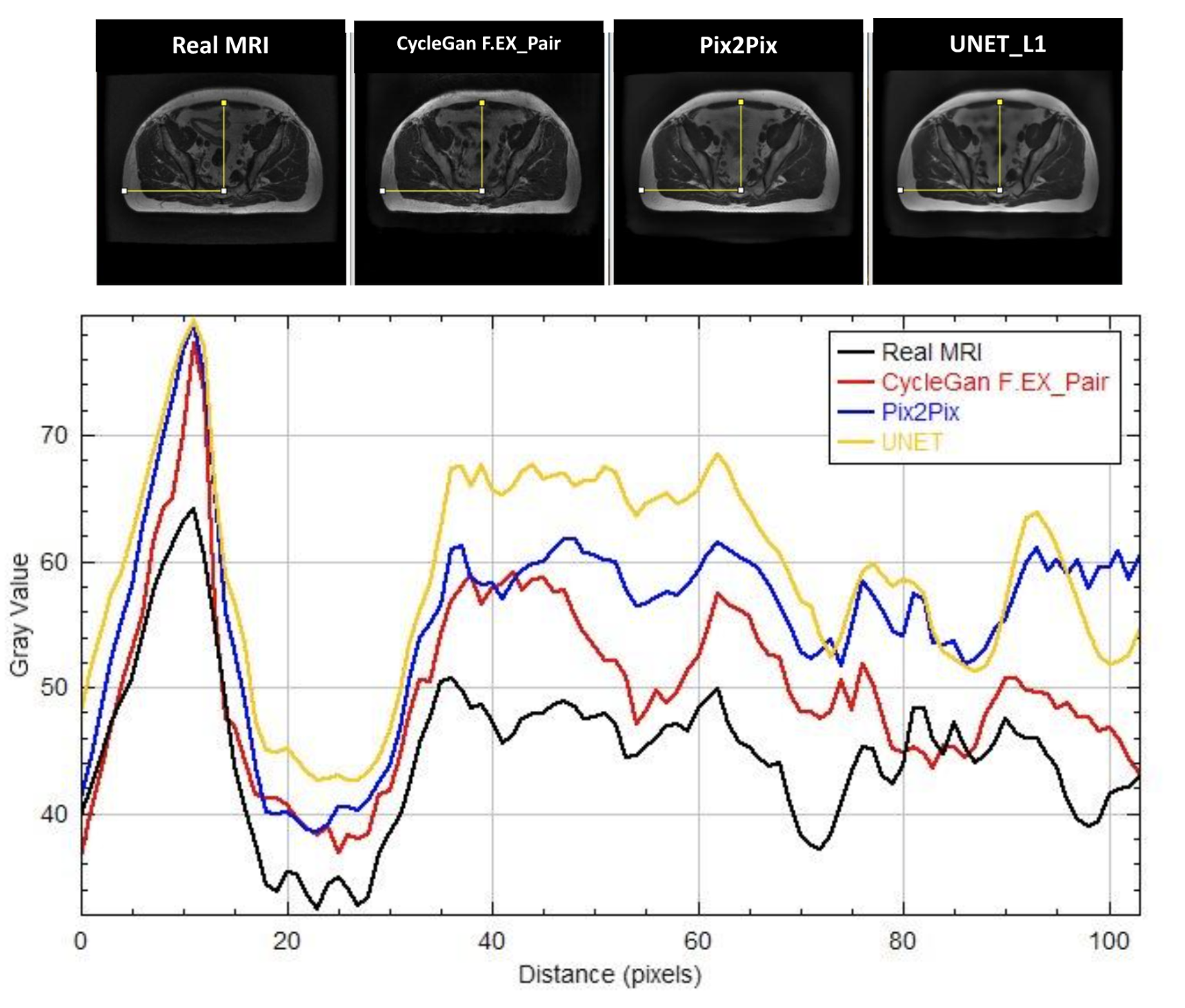}
    \caption{Pixel Intensity Comparison: Models Trained on Paired Images}
    \label{fig:r1}
\end{figure}

While quantitative metrics provided valuable insights, a comprehensive assessment necessitated qualitative evaluation to discern the image quality. The Intensity criterion played a pivotal role; one of the ways to check this criterion is to use a one-dimensional image profile \cite{imagej} indicating that the CycleGAN+FE network synthesized images with brightness levels closely matching real MRI images. As depicted by a selection of a random set of intensity values in Figure~\ref{fig:r1}, it's evident that the CycleGAN+FE Pair outperforms other supervised methods like Pix2Pix and UNet. Even though the training of this network was unsupervised And similar to the CycleGAN+FE Unpair, only paired images have been used to train this network. 

\begin{figure}[!t]
\centering
\subfloat[]{\includegraphics[width=1\linewidth]{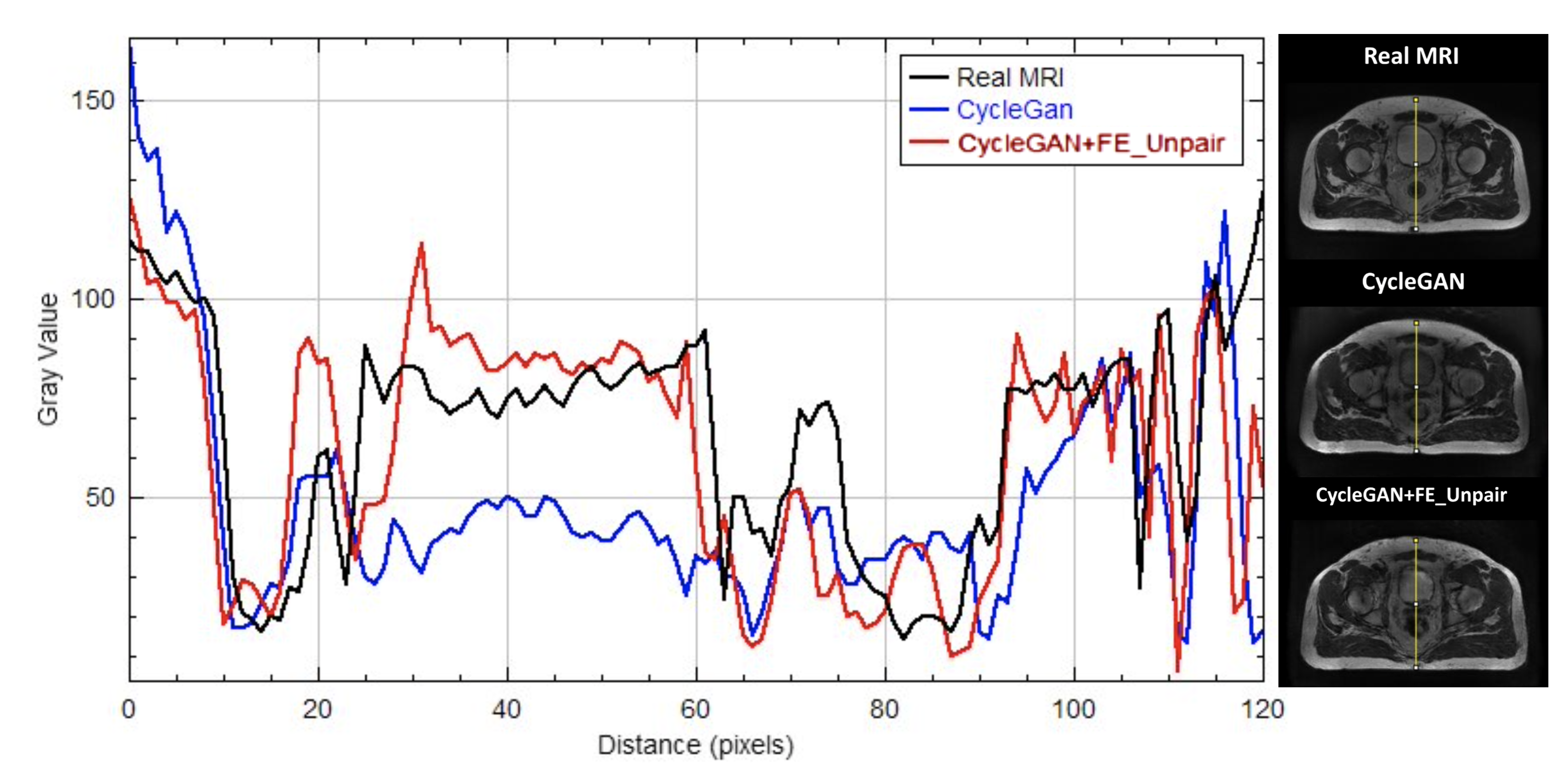}\label{fig_subfig1}}
\hfil
\subfloat[]{\includegraphics[width=1\linewidth]{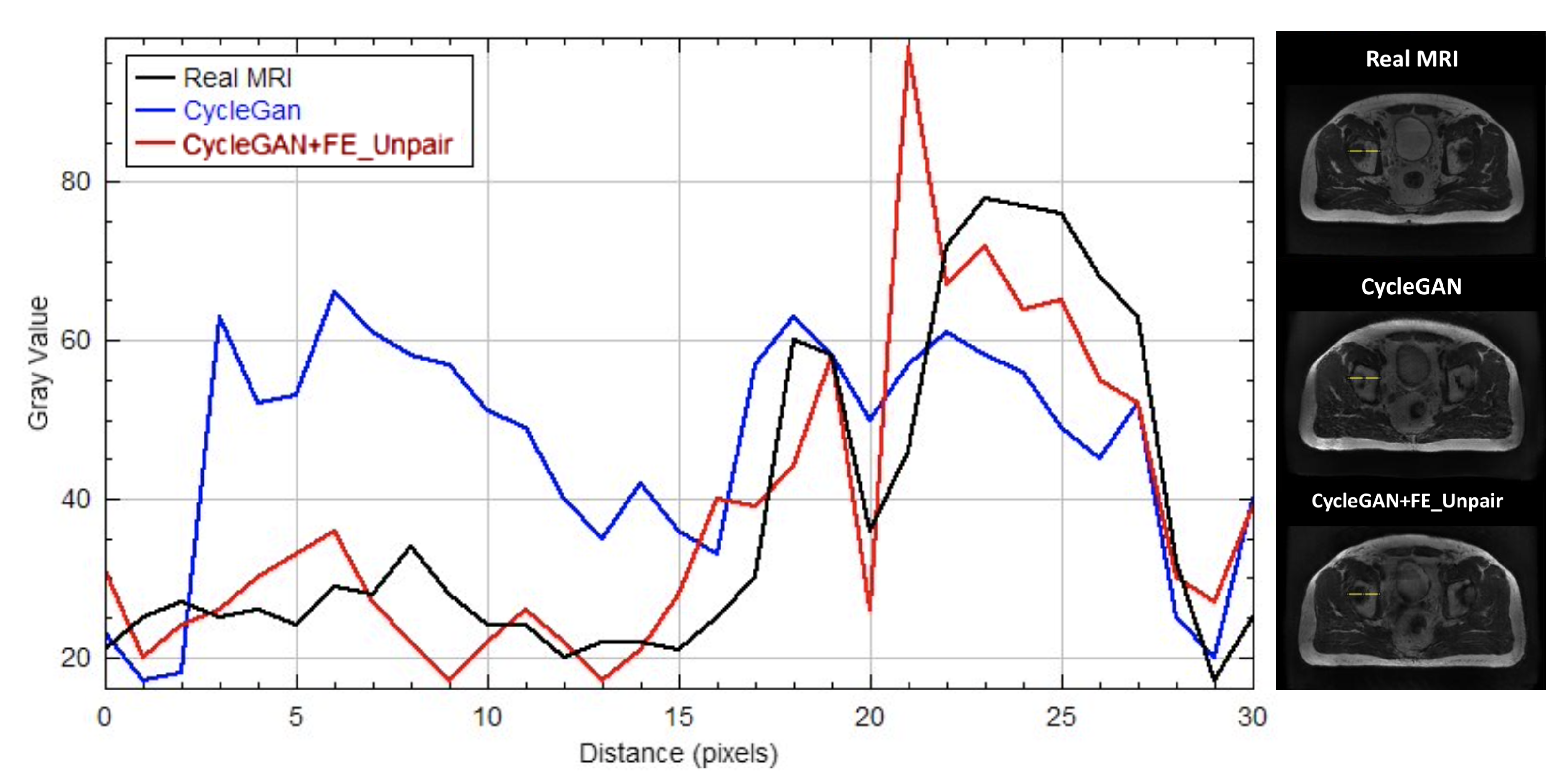}\label{fig_subfig2}}
\caption{Pixel Intensity Comparison: Models Trained on Unpaired Images}
\label{fig:fig_main_1}
\end{figure}

Furthermore, a detailed comparison of the one-dimensional image profile illustrated in Figure~\ref{fig:fig_main_1} for two different synthetic MRI images (a) and (b) shows CycleGAN+FE Unpair significantly outperforms the simple CycleGAN model. This superiority of CycleGAN+FE is particularly evident in its ability to preserve pixel intensities and generate images with enhanced detail.

\begin{figure}[!t]
\centering
\subfloat[]{\includegraphics[width=1\linewidth]{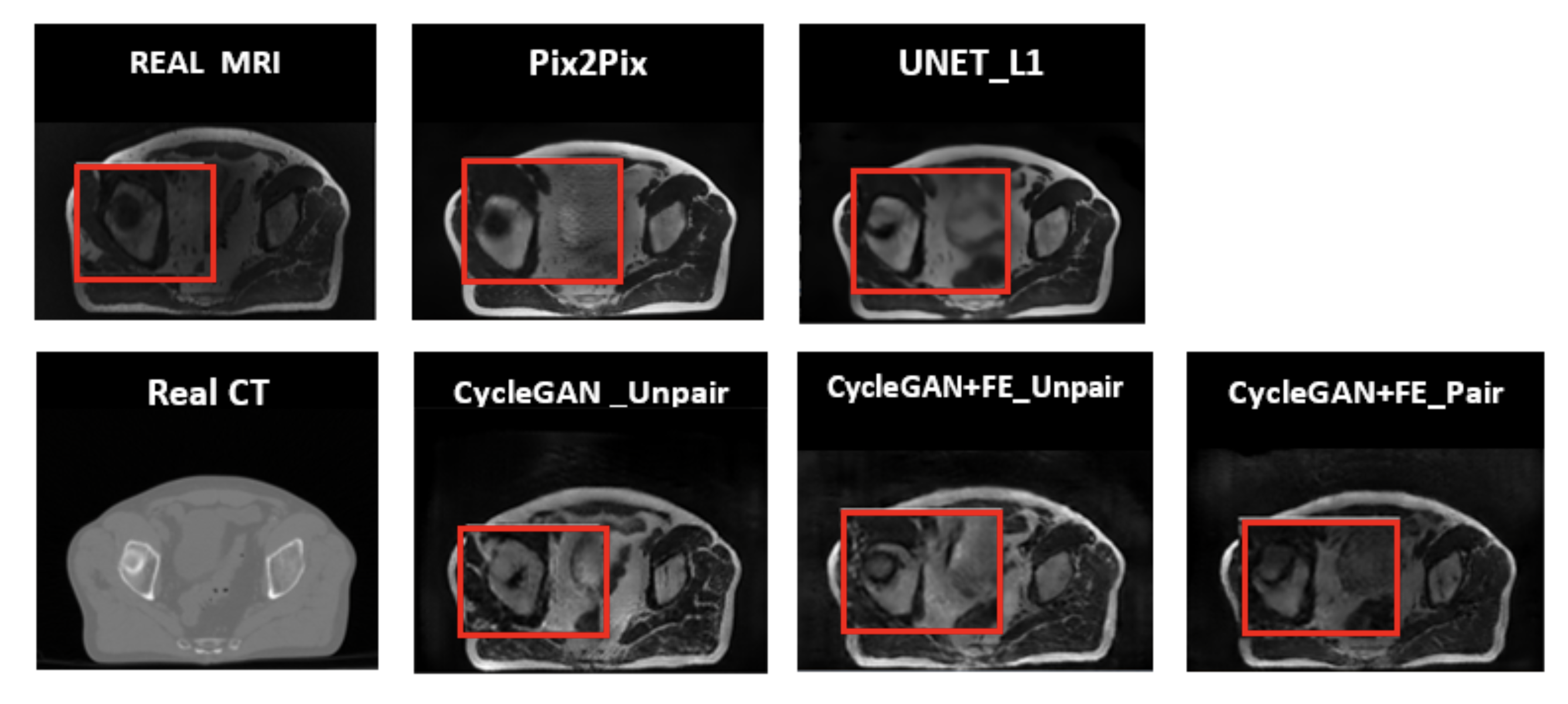}\label{fig:r4}}
\hfil
\subfloat[]{\includegraphics[width=1\linewidth]{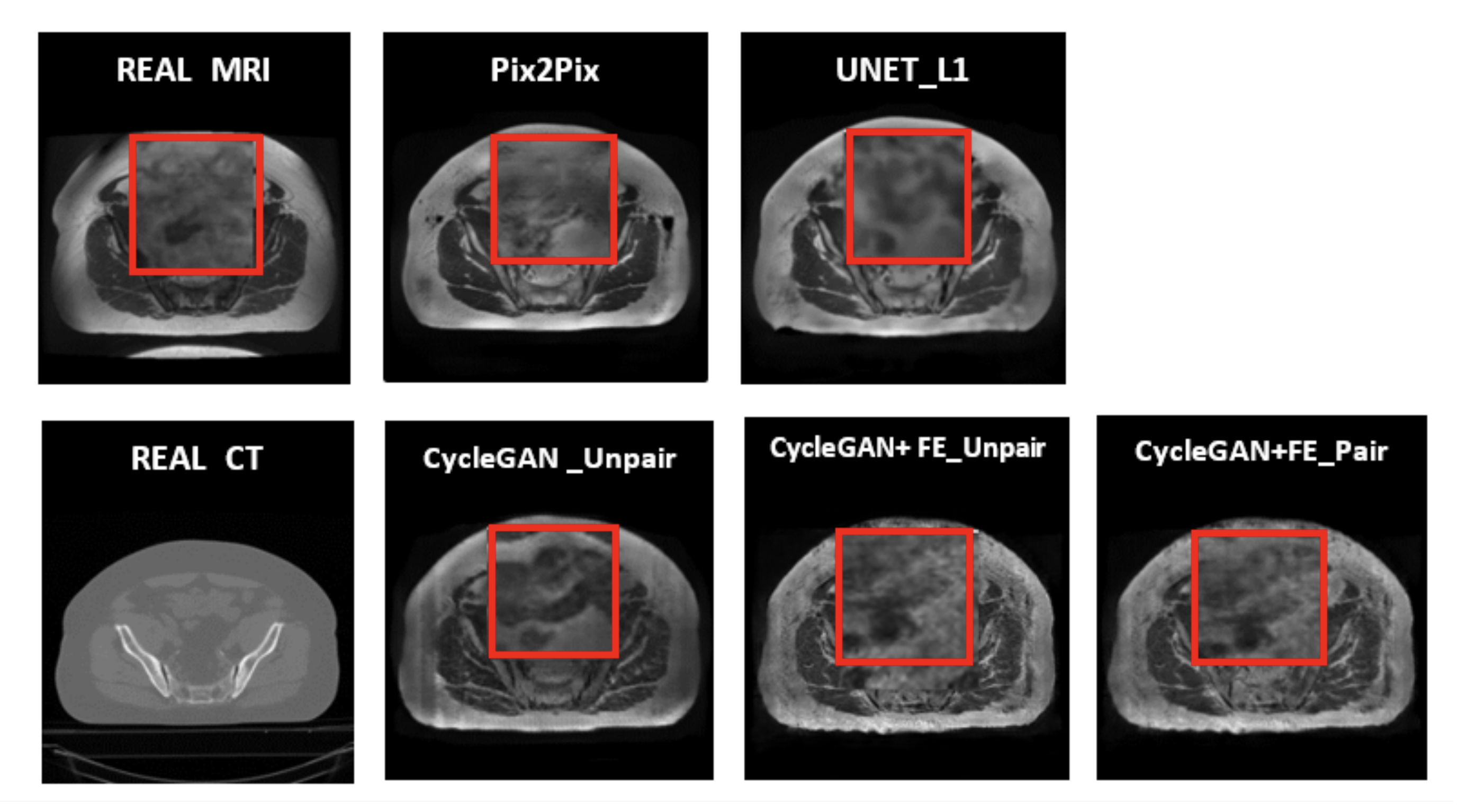}\label{fig:r5}}
\caption{Qualitative Assessment of Synthesized Image Quality}
\label{fig:fig_main_2}
\end{figure}

Upon closer inspection, the enhanced CycleGAN+FE network effectively reduced redundant artifacts common with the simple CycleGAN. In medical imaging, minimizing artifacts is crucial given their significant impact on patient health outcomes. Moreover, the network's capability to preserve tissue details and minimize blurriness further proved its efficiency in producing clinically valuable images. This is evident in Figure~\ref{fig:fig_main_2}, where CycleGAN+FE shows superior detail preservation.

\section{Conclusion and Discussion}
This study has concentrated on enhancing synthetic MRI image reconstruction from CT scans, addressing existing limitations in current deep learning models. Integrating a feature extractor into the CycleGAN network has markedly improved the clarity of image details and the definition of tissue boundaries, outperforming conventional models such as UNet and Pix2Pix. The enhanced CycleGAN+FE network demonstrated superior performance in both Unpaired and Paired data scenarios, as verified through rigorous quantitative and qualitative evaluations.

For future improvements, it is recommended to expand the dataset and incorporate online Data Augmentation to increase diversity in data and better train the network. Shifting focus towards three-dimensional network models could yield more comprehensive reconstructions. Additionally, it is advisable to adjust the network parameters, specifically by gradually reducing the coefficient of the MAE loss function and increasing that of the feature extractor, to attain more realistic images with finer details.

\end{document}